# Tunable optical bistability and tristability of nonlinear graphene-wrapped dielectric nanoparticles

K. Zhang[1], Y. Huang[1,#], Andrey E. Miroshnichenko[2], and L. Gao[1,3,*]

[1]College of Physics, Optoelectronics and Energy of Soochow University, Collaborative Innovation Center of Suzhou Nano Science and Technology, Soochow University, Suzhou 215006, China

[2]Nonlinear Physics Centre, The Australian National University, Canberra ACT 0200, Australia

[3]Jiangsu Key Laboratory of Thin Films, Soochow University, Suzhou 215006, China.

#yanghuang808@163.com

*leigao@suda.edu.cn

## Abstract

Based on full-wave scattering theory with self-consistent mean field approximation, we study the optical multi-stability of graphene-wrapped dielectric nanoparticles. We demonstrate that the optical bistability (OB) of the graphene-wrapped nanoparticle exist in both near-field and far-field spectra, and show the optical multi-stability arising from the contributions of higher-order terms of the incident external field. Moreover, both the optical stable region and the switching threshold values can be tuned by changing either the Fermi level or the size of the nanoparticle. Our results promise the graphene-wrapped dielectric nanoparticle a candidate of multi-state optical switching, optical memories and relevant optoelectronic devices.

## 1. Introduction

Nonlinear optical effects, characterized by nonlinear light-matter interaction, play an important role in modern photonic functionalities [1]. Optical nonlinearities are inherently weak and are superlinearly dependent on the electromagnetic field, one way of enhancing the nonlinear effects is introducing the metal-dielectric structure [2], which supports plasmon resonance in metal–dielectric interface. Surface plasmon, arising from coherent oscillations of conduction electrons near the surface of noble-metal structures [2], results in strong confinement and enhancement of local electromagnetic field, boosting the nonlinear optical effects of the metal-dielectric structures.

On the other hand, Optical bistability is one remarkable feature of nonlinear optical effects [1]. It suggests a new way of manipulating the light by light, where a nonlinear optical system exhibits two distinguished stable excited states for a single input intensity [3]. Optical structures with such property can be the candidate for all-optical switching, optical transistor and optical memory [4, 5]. One way of analyzing the bistability is to explore the optical Kerr effect [1], a nonlinear phenomenon where the light modulates the material's refractive index, and several works have already been done [6, 7].

Nevertheless, due to intrinsic Ohmic losses [2], plasmonic field enhancement is fundamentally limited even in noble metals, which constrains the functionality of some of metamaterials and transformation optical devices. Graphene, with extremely large electron mobilities [8] and concomitant extraordinary plasmonic field enhancements, has attracted significant interest in the plasmonic community [9-12], suggesting a platform for metamaterials and transformation optical devices [13]. Moreover, graphene's intrinsic nonlinear properties have been proved both theoretically [14–16] and experimentally [17-19]. Combine the nonlinearity and plasmonic excitation properties together, a body of research is rapidly emerging at the crossroad of nonlinear plasmonics and graphene [20-24]. Besides, there has also been some theoretical interest in optical bistability involving the nonlinear effect of graphene in a free-standing graphene sheet [25], in 2D graphene nanoribbon [13] and in sandwiched structure [26]. Unlike the 1D or 2D system mentioned above, we present a 3D structure, where a dielectric nanosphere is wrapped by graphene, to study the unstable behavior.

Graphene-wrapped structures have been studied both theoretically [12, 27] and experimentally [28-30], where the flexible core-diameters range from tens nanometers to several micrometers, however, few focus on the nonlinear effects of such composite structures. More recently, in the quasistatic limit, we theoretically study the effective third-order nonlinear response and optical bistability of the three-dimensional nanocomposites, containing nonlinear graphene wrapped dielectric nanoparticles embedded in the host medium [31]. In this paper, we investigate the optical bistable (tristable) behavior in the near-field and far-field spectra from graphene-wrapped spheres by generalizing linear full-wave scattering theory to nonlinear theory [32]. In conjunction with the self-consistent mean-field method [7], we demonstrate that the threshold values of both OB and OT are tunable either by varying the particle size or changing the Fermi energies. Our results are useful in the design of multi-state optical switching, which has potential applications in optical communications and computing.

## 2. Theoretical models and methods

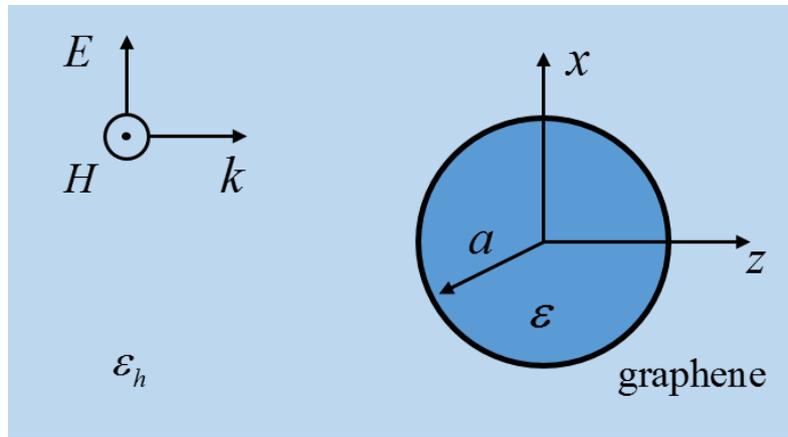

**Fig.1. Geometric model of graphene-wrapped dielectric particle embedded in the dielectric host. The incident plane wave is polarized in x-direction and propagates along z-direction**.

Let's first consider the structure consisting of the monolayer graphene-wrapped dielectric spheres of radius $a$ and permittivity $\varepsilon$, embedded in a host medium with $\varepsilon_h$. Importantly, such graphene-wrapped nanoparticle can be now fabricated experimentally [28-30].

## A. Full-wave theory for linear graphene-coated nanosphere

The incident electric fields upon the graphene-wrapped dielectric nanoparticle has the form

$$\mathbf{E}_{in} = \hat{x}\mathrm{E}_0 e^{ikz} \cdot e^{-i\omega t}, \qquad (1)$$

Where $k = k_0\sqrt{\varepsilon_h}$ denotes the wave vector in the host medium, and $k_0 = \omega\sqrt{\varepsilon_0\mu_0}$ is the one in vacuum. Compared with the size of the dielectric nanoparticle, the monolayer graphene is only one atom thick and it can be considered as an extremely thin conducting shell with linear conductivity $\sigma_g$ [33]. Hence we employ the following boundary conditions,

$$\hat{n} \times (\mathbf{H}_i + \mathbf{H}_s - \mathbf{H}_c) = \mathbf{J} \qquad (2)$$

$$\hat{n} \cdot (\mathbf{E}_i + \mathbf{E}_s - \mathbf{E}_c) = 0 \qquad (3)$$

Where, $\mathbf{H}_i$, $\mathbf{E}_i$, $\mathbf{H}_s$, $\mathbf{E}_s$, $\mathbf{H}_c$, and $\mathbf{E}_c$ are the incident, scattered, and internal magnetic and electric fields of the dielectric sphere, respectively. $\mathbf{J} = \sigma_g \mathbf{E}_t$ is the surface current density induced by the tangential component of the electric field $\mathbf{E}_t$.

According to Mie scattering theory [32], the general solutions for the local electromagnetic field can be written as follows,

$$\begin{aligned}
\mathbf{E}_c &= \sum_{n=1}^{\infty} E_n \left( c_n \mathbf{M}_{o1n}^{(1)} - i d_n \mathbf{N}_{e1n}^{(1)} \right) \\
\mathbf{H}_c &= (-k_1 / \omega\mu) \sum_{n=1}^{\infty} E_n \left( d_n \mathbf{M}_{e1n}^{(1)} + i c_n \mathbf{N}_{o1n}^{(1)} \right) \\
\mathbf{E}_{out} &= \mathbf{E}_i + \mathbf{E}_s = \sum_{n=1}^{\infty} E_n \left( \mathbf{M}_{o1n}^{(1)} - i\mathbf{N}_{e1n}^{(1)} \right) + \sum_{n=1}^{\infty} E_n \left( i a_n \mathbf{N}_{e1n}^{(3)} - b_n \mathbf{M}_{o1n}^{(3)} \right) \\
\mathbf{H}_{out} &= \mathbf{H}_i + \mathbf{H}_s = (-k / \omega\mu) \left( \sum_{n=1}^{\infty} E_n \left( \mathbf{M}_{e1n}^{(1)} + i\mathbf{N}_{o1n}^{(1)} \right) - \sum_{n=1}^{\infty} E_n \left( i b_n \mathbf{N}_{e1n}^{(3)} + a_n \mathbf{M}_{o1n}^{(3)} \right) \right)
\end{aligned} \tag{4}$$

where $E_n = i^n E_0 (2n+1)/n(n+1)$, $\mathbf{M}_{\sigma 1n}^{(1),(3)} = \nabla \times \left[ \mathbf{r} \dfrac{j_n}{h_n}(kr) Y_{\sigma 1n}(\theta, \varphi) \right]$, $\mathbf{N} = \dfrac{1}{k}\nabla \times \mathbf{M}$.

Here $\mathbf{M}$ and $\mathbf{N}$ are the vector spherical harmonics, and the upper indices (1) and (3) indicate the use of the spherical Bessel function $j_n$ and the first-order spherical Hankel function $h_n$, respectively, and

$$Y_{\substack{e\\o}1n}(\theta, \varphi) = P_n^{(1)}(\cos\theta) \begin{matrix}\cos\\ \sin\end{matrix}(\varphi), \tag{5}$$

where $P_n^{(1)}$ are the associated Legendre functions. In addition, $k_1 = k_0 \sqrt{\varepsilon}$ is the wave number inside the sphere. $a_n, b_n, c_n$ and $d_n$ are the five unknown coefficients to be determined. Applying boundary conditions in Eq. (2) and (3), we will achieve the following coefficients, which agrees with solutions in [12],

$$\begin{aligned}
a_n &= \frac{\psi_n(x)\psi_n'(mx) - m\psi_n'(x)\psi_n(mx) - i\sigma_g \alpha \psi_n'(x)\psi_n'(mx)}{\xi_n(x)\psi_n'(mx) - m\xi_n'(x)\psi_n(mx) - i\sigma_g \alpha \xi_n'(x)\psi_n'(mx)} \\
b_n &= \frac{\psi_n(mx)\psi_n'(x) - m\psi_n'(mx)\psi_n(x) + i\sigma_g \alpha \psi_n(x)\psi_n(mx)}{\psi_n(mx)\xi_n'(x) - m\psi_n'(mx)\xi_n(x) + i\sigma_g \alpha \xi_n(x)\psi_n(mx)} \\
c_n &= \frac{m\psi_n(x)\xi_n'(x) - m\psi_n'(x)\xi_n(x)}{\psi_n(mx)\xi_n'(x) - m\psi_n'(mx)\xi_n(x) + i\sigma_g \alpha \xi_n(x)\psi_n(mx)} \\
d_n &= \frac{m\psi_n'(x)\xi_n(x) - m\psi_n(x)\xi_n'(x)}{\xi_n(x)\psi_n'(mx) - m\xi_n'(x)\psi_n(mx) - i\sigma_g \alpha \xi_n'(x)\psi_n'(mx)}
\end{aligned} \tag{6}$$

where $\psi_n(x) = x j_n(x)$, $\xi_n(x) = x h_n(x)$, $\alpha = \sqrt{\mu_0 / \varepsilon_0 \varepsilon_h}$. $m = k_1 / k = \sqrt{\varepsilon / \varepsilon_h}$ is defined as relative refractive index, and $x = ka$ is the size parameter. Note that when $\sigma_g = 0$, all the coefficients shown in Eq. (7) reduce to those in the case of a bare spherical dielectric particle.

The distribution of the local electric field in the graphene thin layer can be obtained with Eq. (4) with $r = a$, and the spatial average of the square of the linear local electric field within the graphene, which will be generalized in the following subsection, can be written as,

$$\left\langle |\mathbf{E}|^2 \right\rangle_{lin,g} = N |\mathrm{E}_0|^2 \qquad (7)$$

with $N = \dfrac{1}{4\pi} \int_0^{2\pi} \int_0^{\pi} \sum_{n=1}^{\infty} \left( \dfrac{2n+1}{n(n+1)} \right)^2 \left| c_n \mathbf{M}_{o1n}^{(1)} - i d_n \mathbf{N}_{e1n}^{(1)} \right|^2 \sin\theta d\theta d\varphi$.

In addition, we define the linear scattering cross section efficiency as [32],

$$Q_{\mathrm{sca}} = \frac{2}{(ka)^2} \sum_{n=1}^{\infty} (2n+1)\left(|a_n|^2 + |b_n|^2\right). \qquad (8)$$

## B. Nonlinear theory for nonlinear graphene-coated nanosphere

In general, the surface conductivity of the graphene is field-dependent and nonlinear, and we introduce the simplified version within the random-phase approximation [27],

$$\tilde{\sigma}_g = \sigma_g + \sigma_3 |\mathbf{E}|^2, \qquad (9)$$

in which the linear term $\sigma_g = \sigma_{\mathrm{intra}} + \sigma_{\mathrm{inter}}$, and $\sigma_{\mathrm{intra}}$, $\sigma_{\mathrm{inter}}$ are the intraband and interband terms which have the following forms [27],

$$\begin{aligned}
\sigma_{\mathrm{intra}} &= \frac{ie^2 k_B T}{\pi \hbar^2 (\omega + i/\tau)} \left[ \frac{E_F}{k_B T} + 2\ln\left( e^{-\frac{E_F}{k_B T}} + 1 \right) \right] \\
\sigma_{\mathrm{inter}} &= \frac{ie^2}{4\pi \hbar} \ln \left| \frac{2E_F - (\omega + i\tau^{-1})\hbar}{2E_F + (\omega + i\tau^{-1})\hbar} \right|
\end{aligned} \qquad (10)$$

where $e, \hbar, k_B, E_F, \tau$ and $k_1 = k_0 \sqrt{\varepsilon}$ are electron charge, reduced Plank constant, Boltzmann constant, Fermi energy, electron-phonon relaxation time and temperature respectively. The Fermi energy $E_F = \hbar v_F \sqrt{\pi n_{2D}}$ can be electrically controlled by an applied gate voltage due to the strong dependence of the carrier density $n_{2D}$ on the gate voltage, and the relaxation time is determined by the carrier mobility $m_c$ as $\tau = m_c E_F / e v_F^2$ [30]. $v_F$ is the Fermi velocity of electrons.

For photon energy $\hbar\omega$ far less than the Fermi energy $E_F$, the interband transitions in graphene is negligible compared with the intraband part. Therefore, in the THz range graphene is well described by the Drude-like surface conductivity $\sigma_{\mathrm{intra}}$. And $E_F \gg k_B T$ in the room temperature ($T = 300\mathrm{K}$), hence we have the simplified form of graphene linear conductivity $\sigma_g$ and third-order nonlinear surface conductivity $\sigma_3$ [27, 28],

$$\sigma_g = \frac{ie^2 E_F}{\pi \hbar^2 (\omega + i/\tau)}, \quad \sigma_3 = -i \frac{9 e^4 v_F^2}{8\pi E_F \hbar^2 \omega^3} . \qquad (11)$$

As one knows, when the nonlinear conductivity of the coated graphene is taken into account, the local fields inside the dielectric nanoparticle may become inhomogeneous, which might make it difficult to determine the filed-dependent conductivity [Eq. (10)] at each point inside the graphene layer. As a consequence, the local fields within the graphene-wrapped nanoparticle are not solved exactly. To give a simple and efficient way, we alternatively introduce the mean-field approximation [6], in which the nonlinear local field $|E|^2$ within atomically thin graphene layer is replaced by the mean one $\langle |E|^2 \rangle_{non,g}$. Therefore Eq. (10) turns to

$$\tilde{\sigma}_g \approx \sigma_g + \sigma_3 \langle |\mathbf{E}|^2 \rangle_{non,g}. \tag{12}$$

On the other hand, Eq. (8) can be modified as,

$$\langle |\mathbf{E}|^2 \rangle_{non,g} = \tilde{N} |\mathbf{E}_0|^2 \tag{13}$$

with $\tilde{N} = \frac{1}{4\pi} \int_0^{2\pi} \int_0^{\pi} \sum_{n=1}^{\infty} (-1)^n \left( \frac{2n+1}{n(n+1)} \right)^2 \left| \tilde{c}_n \mathbf{M}_{o1n}^{(1)} - i\tilde{d}_n \mathbf{N}_{e1n}^{(1)} \right|^2 \sin\theta d\theta d\varphi$. Here we mention that $\tilde{c}_n$ and $\tilde{d}_n$ have the same form as $c_n$ and $d_n$ but with $\sigma_g$ in $c_n$ and $d_n$ replaced by the field-dependent conductivity $\tilde{\sigma}_g \approx \sigma_g + \sigma_3 \langle |\mathbf{E}|^2 \rangle_{non,g}$. Hence, $\tilde{N}$ in above equation is dependent on $\langle |E_c|^2 \rangle$ and Eq. (14) is a self-consistent equation for $\langle |\mathbf{E}|^2 \rangle_{non,g}$. As a consequence, one can obtain the relation between local field $\langle |\mathbf{E}|^2 \rangle_{non,g}$ and external field $|\mathbf{E}_0|^2$ by directly solving Eq. (14) in a self-consistent manner, and hence one may achieve the optical bistable behavior for the near field.

The nonlinear scattering efficiency $\tilde{Q}_{sca}$ of the graphene-wrapped nanoparticle can also be generalized as,

$$\tilde{Q}_{sca} = \frac{2}{(ka)^2} \sum_{n=1}^{\infty} (2n+1) \left( |\tilde{a}_n|^2 + |\tilde{b}_n|^2 \right), \tag{14}$$

again, $\tilde{a}_n$ and $\tilde{b}_n$ have the same form as $a_n$ and $b_n$ but with $\sigma_g$ in $a_n$ and $b_n$ replaced by the field-dependent conductivity $\tilde{\sigma}_g \approx \sigma_g + \sigma_3 \langle |\mathbf{E}|^2 \rangle_{non,g}$.

## 3. Numerical results

We are now in a position to provide some numerical results. To start with, let's first investigate the scattering property of the monolayer graphene-wrapped dielectric sphere in the linear case, by neglecting the field-dependent term of the surface conductivity in Eq. (9). In the THz frequencies, the imaginary part of the surface conductivity of the graphene is positive and indicates a "metallic" type nature which can support surface plasma [33]. Therefore, this graphene layer acts as a very thin "metallic" shell and leads to the resonant peaks in the scattering efficiency spectra as shown in Fig. 2.

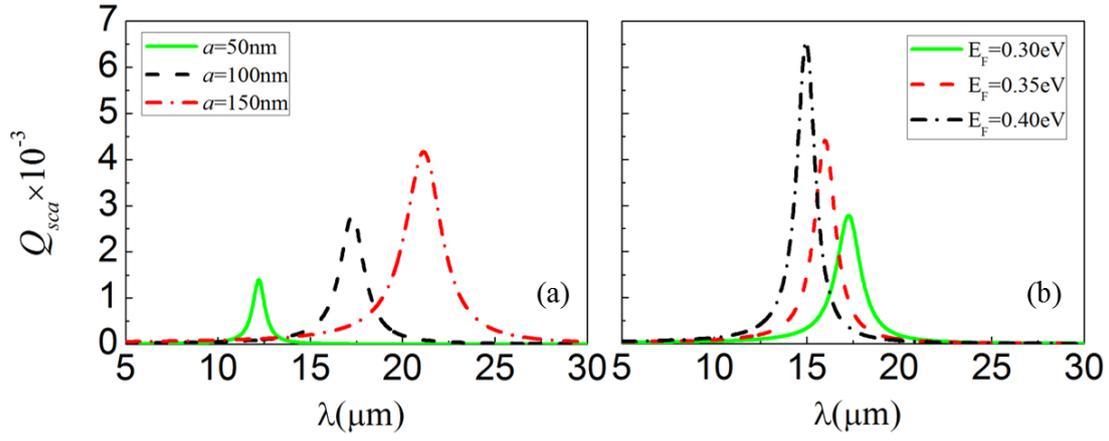

**Fig.2. Linear scattering efficiency of the monolayer graphene-wrapped dielectric sphere as a function of incident wavelength, with** (a) **different sphere radius at Fermi energy** $E_F = 0.3\text{eV}$; **and** (b) **with different Fermi energy at a fixed sphere radius** $a = 100\text{nm}$. **Other parameters are** $\varepsilon = \varepsilon_h = 2.25$ **and** $\tau = 0.1\text{ps}$, **respectively.**

Unlike metallic nanoparticle with small size, the graphene-wrapped dielectric nanoparticle has tunable plasmonic resonances associated with the particle size in its linear scattering efficiency spectra. Fig. 2(a) indicates that when the radius of the nanoparticle is increased from 50nm to 100nm, the resonant peaks are red-shifted, accompanied with enhanced peaks. In the present model, although the radius of the nanoparticle is much smaller than the incident wavelength, with the plasmonic nature of the graphene layer in the THz frequencies, we can still achieve size-dependent resonances, beyond the Rayleigh approximation. For a metallic sphere, however, the plasmonic resonances are almost independent on the size when the incident wavelength is much longer. In addition, besides the radius of the nanoparticle, different Fermi levels of graphene can also modify the surface conductivity, hence lead to the tunable resonances as well. Fig. 2(b) shows the dependence of the resonant peak on the Fermi level, it clearly indicates that higher Fermi level will generally result in the blue-shift of the resonant wavelength and enhanced peak. In a different structure, our results qualitatively coincide with those in the graphene ribbons described by other group [10].

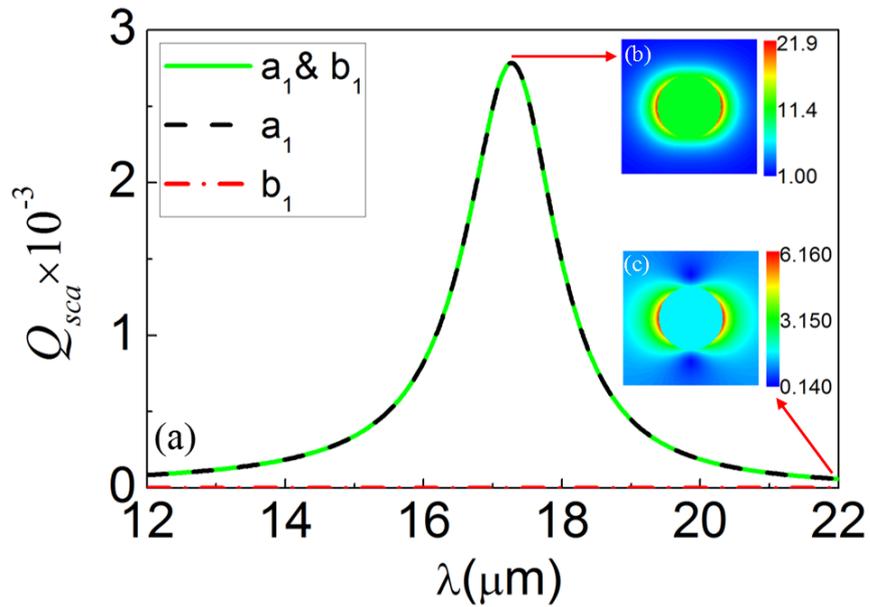

**Fig.3.** (a) **Dependence of** $Q_{sca}$ **on incident wavelength with** $a = 100\text{nm}$, $E_F = 0.3\text{eV}$, $\varepsilon = \varepsilon_h = 2.25$ **and** $\tau = 0.1\text{ps}$. **And correspondent distributions of the electric fields inside and outside the sphere for** (b) $a = 100\text{nm}$, $\lambda = 17.3\mu m$; (c) $a = 100\text{nm}$, $\lambda = 22\mu m$.

To observe strong nonlinear effects in the graphene described by a field-dependent surface conductivity, i.e. Eq. (9), strong field intensity is required due to the fact that third-order nonlinear coefficient $\sigma_3$ generally has very small value. Localized surface plasmon resonance would enhance the local field so that it can be further used to boost the naturally weak nonlinear effects in this graphene-wrapped nanoparticle. Along this line, we would like to take one step forward to investigate the spatial local fields near the resonant wavelength. In Fig. 3, we plot the $Q_{sca}$ with a fixed radius $a = 100$nm. It is obvious that the field-enhancement $\lambda = 17.3\mu m$ is larger than that of $\lambda = 22\mu m$, which is mainly due to the surface plasmon resonant contribution. Note that the scattering efficiency depicted by Eq. (8) depends on the Mie coefficients for TM ($a_n$) and TE ($b_n$) modes, while the resonance here totally results from the electric dipole resonance, which can be proved by comparing the green (solid) line with the black (dashed) line and the red (dash-dotted) line in Fig. 3 (a).

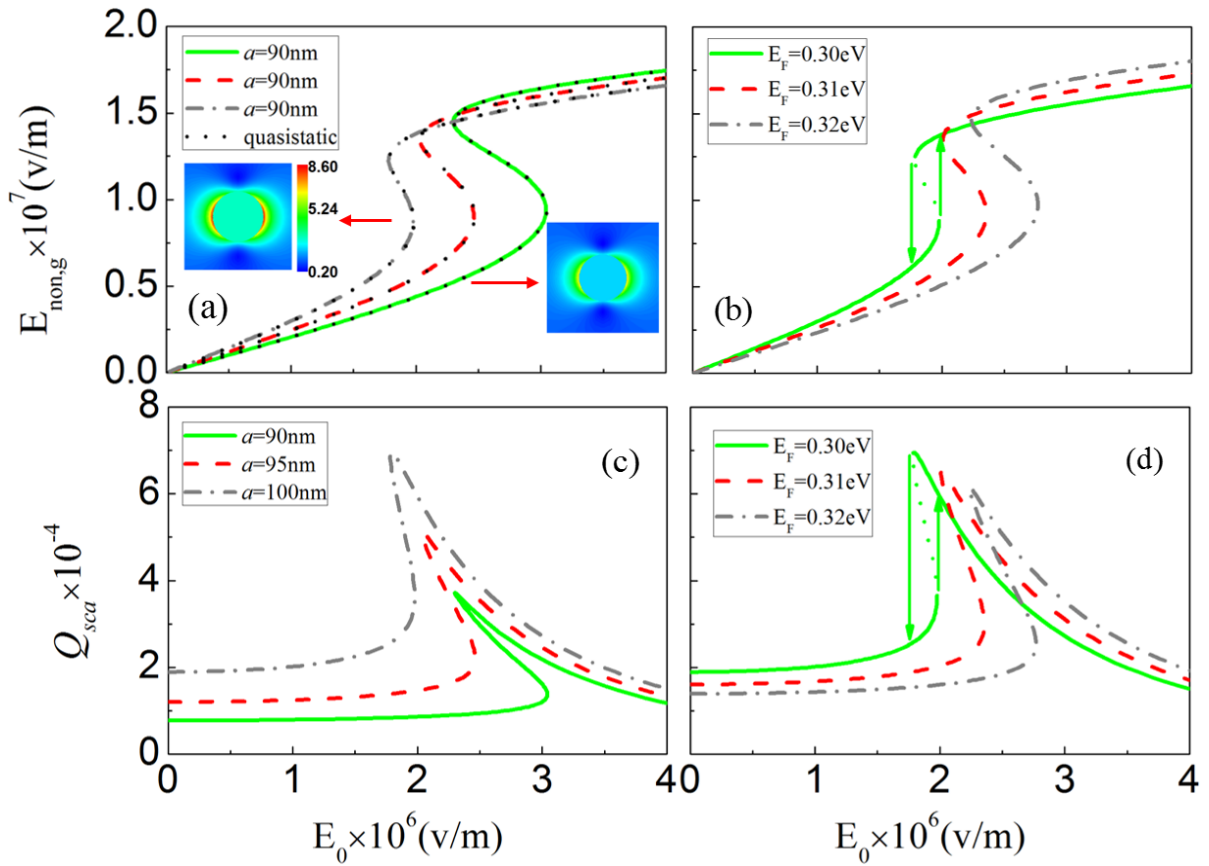

**Fig.4.** The average local field $E_{non,g}$ as a function of the external applied field $E_0$ for (a) same Fermi energy $E_F = 0.3$eV with different radii. Solid lines represent the full-wave scattering results and dotted lines indicates the quasi-static limit. The inserts show the linear near-field distribution. (b) same radius $a = 100$nm with different Fermi energy. (c) and (d) are nonlinear scattering efficiency versus $E_0$. Other parameters are $\varepsilon = \varepsilon_h = 2.25$, $\tau = 0.1$ps and $\lambda = 20\mu m$.

In what follows, we'd like to study the optical switching effects in such nonlinear nanosphere near the surface plasmon resonant wavelengths. We plot the average of the local field $E_{non,g} \equiv \sqrt{\langle |E|^2 \rangle_{non,g}}$ within the graphene-wrapped nanoparticle as a function of the applied external field for several sizes and different Fermi levels in Fig. 4 (a) and Fig. 4(b). From Fig. 4(a) and (b), it shows that the average of the local-field within the nanoparticles firstly increases with increasing the applied field, and it jumps discontinuously to the upper branch when the applied field reaches the switching-up threshold field. Further increasing the external incident field leads to monotonic increase of the local-field. On the other hand, the continuous decrease of the applied field will lead to the discontinuous jump of the local field from the upper branch to the lower branch at the switching-down threshold field. This OB behavior in the near field shows a potential application in nano-switches and nano-memories. Moreover, larger size of the nanoparticle will generally result in a lower bistable switching-up threshold field accompanied by a narrower bistable region. This can be understood as follows: the magnitude of the linear local field within the graphene thin layer is indeed enhanced as increasing the radii of the nanoparticle [see the insert of Fig. 4(a)], resulting in the decrease of the threshold field. The results given by the quasistatic approximation [31] are shown in Fig. 4(a) (dotted line) as well to make comparison. It exhibits the results achieved by using the quasistatic approximation matches quite well with our nonlinear full-wave theory, indicating that both the quasistatic approximation and the full wave theory in the study of nonlinear OB are accurate within the dimension of the nano-device. Fig. 4(b) illustrates the influence of Fermi level on bistable near-field. In this case, the threshold values differ mainly because the Fermi energy influences the nonlinear coefficient of graphene conductivity, i.e. Eq. (11), which in turn affects the average local field. Different from the size effect, increasing the Fermi level of the graphene will make the switching-up threshold become higher and lead to a broader bistable region. Note that this graphene-wrapped dielectric nanoparticle provides us one possible way to realize nonlinear optical switch devices, whose switching threshold field can be tunable by changing the Fermi level.

Next, the dependence of the far field scattering property of this graphene-wrapped nanoparticle is investigated. By substituting Eq. (13) into field-dependent scattering efficiency in Eq. (14), one yields the nonlinear scattering efficiency as the function of the applied external field, as shown in Fig. 4 (c) and (d). Again, nonlinear scattering efficiency reveals optical bistable behavior unlike conventional hysteretic loops, the upper branch of these bistable curves exhibit very abrupt variation compared with the lower one, and the maximal scattering efficiency before dropping at switching-down threshold is extremely high. This novel property may give rise to some potential applications in nonlinear optical sensors.

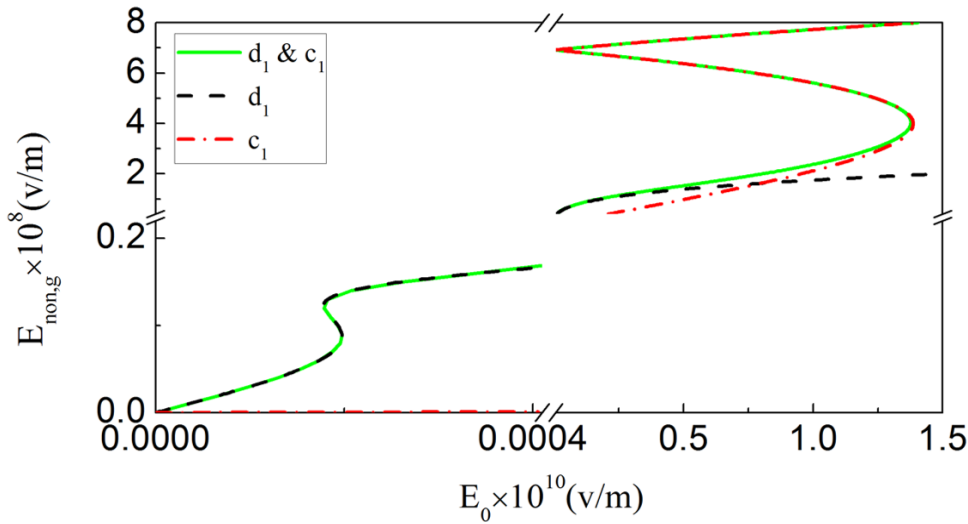

**Fig.5. Contributions from electric dipole (dotted line) and magnetic dipole (dash-dotted line) to the bistable behavior with the increasing of applied field $E_0$.**

As we further increase the magnitude of the applied field, to our interest, we find another hysteresis curve, which is mainly because that, like the nonlinear scattering efficiency, the nonlinear local field also depends on the Mie coefficients for TM ($d_n$) and TE ($c_n$) modes as shown by Eq. (13). Combining Fig. 4 (a) with Fig. 5, we conclude that the former bistable behavior origins totally from the electric dipole resonance. However, much larger than the resonant field, the increased applied field results in the excitation of the magnetic dipole contribution, boosting the nonlinearity of graphene, hence, the latter hysteresis curve can be expected.

Before, researches were conducted by just considering the first order term ($n=1$), since the sizes we adopted above are much smaller than the incident wavelengths, and the applied field is low, the theoretical results on OB based on the nonlinear full-wave theory are in good with those in the quasistatic limit. However, things changed when we increase the applied field, because it is unable to excite the magnetic dipole contribution in the quasistatic limit. Hence we take on more step to see how the higher orders affect the nonlinear properties of the proposed graphene-wrapped nanoparticle with relevantly larger size in the near-field spectra. Achieved results are shown below, optical multistates are observed, and one should note that there are no electric dipole or even multipole resonance caused optical behaviors when we adopt the parameters $a=1\mu m$ and $\lambda=40\mu m$, which can be concluded from Fig.4 (a) that the unstable region would narrow and eventually disappear if we increase the size of the nanoparticle. So the hysteresis curves here mainly result from the magnetic multipole contributions excited by high incident field.

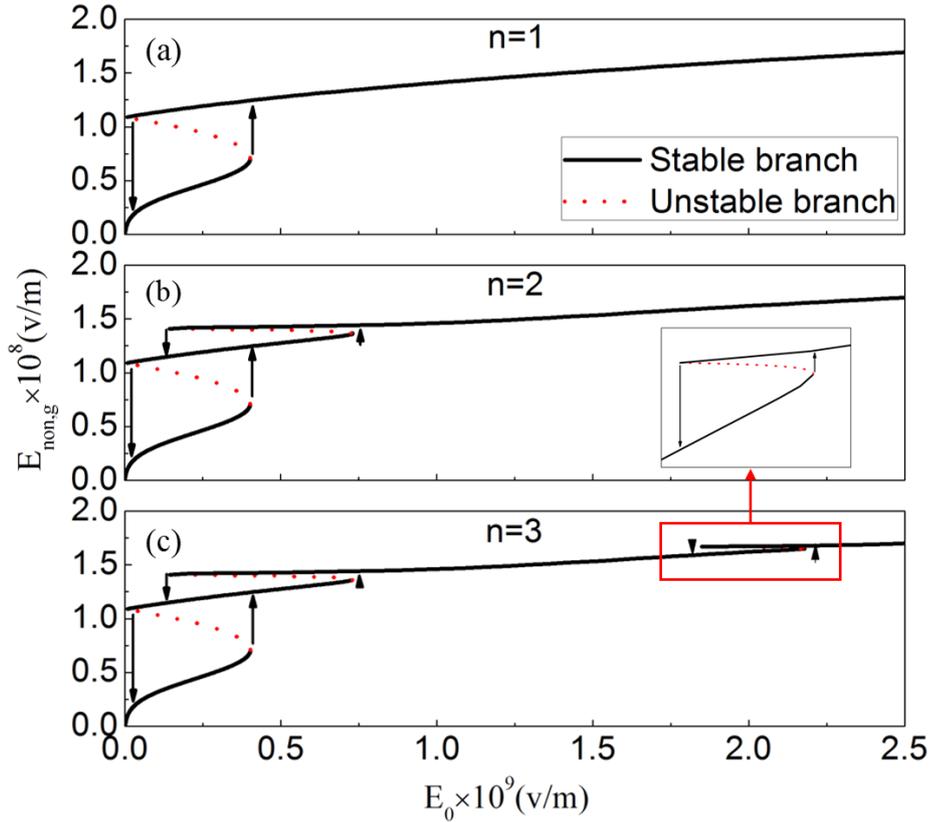

**Fig.6. Three typical behaviors in nonlinear composites for $a=1\mu m$, and $\lambda=40\mu m$, with considering the nth order of incident TM wave:** $(a)$ **the first order;** $(b)$ **the first and second orders; and** $(c)$ **the first, second and third orders.**

Taking a close look at Fig. 6, it is found that once the higher order terms are taken into account the nonlinear curves will become more complicated. For instance, magnetic dipole order together with quadrupole term ($n=2$) will contribute to the optical tristability (OT)[6], which has one more stable branch from the magnetic quadrupole term compared to the magnetic dipole term caused one [see Fig.6 (a) and (b)]. OT curve in Fig. 6 (b), reveals the following nonlinear process, when the applied external field $E_0$ starts to increase over the first upper threshold field, the discontinuous jump of the local field takes place from the lower branch to the middle branch; as $E_0$ further increases up to the second upper threshold field, we find the other discontinuous jump from the middle branch to the upper branch. On the contrary, if one decreases $E_0$ the average local field will first jump to the middle branch before jump to the lower branch. The difference between the OB and OT is that for a given $E_0$ the average field has three real roots within one electric field domain, hence the desired OB. However, it has five real roots in one incident field region, hence the desired OT. Besides the magnetic dipole and quadrupole terms, once the magnetic octupole term

($n=3$) is included another bistable curve exists at higher $E_0$ region in Fig. 6 (c). Therefore, one might achieve more functionality about optical switching in this proposed nano-device.

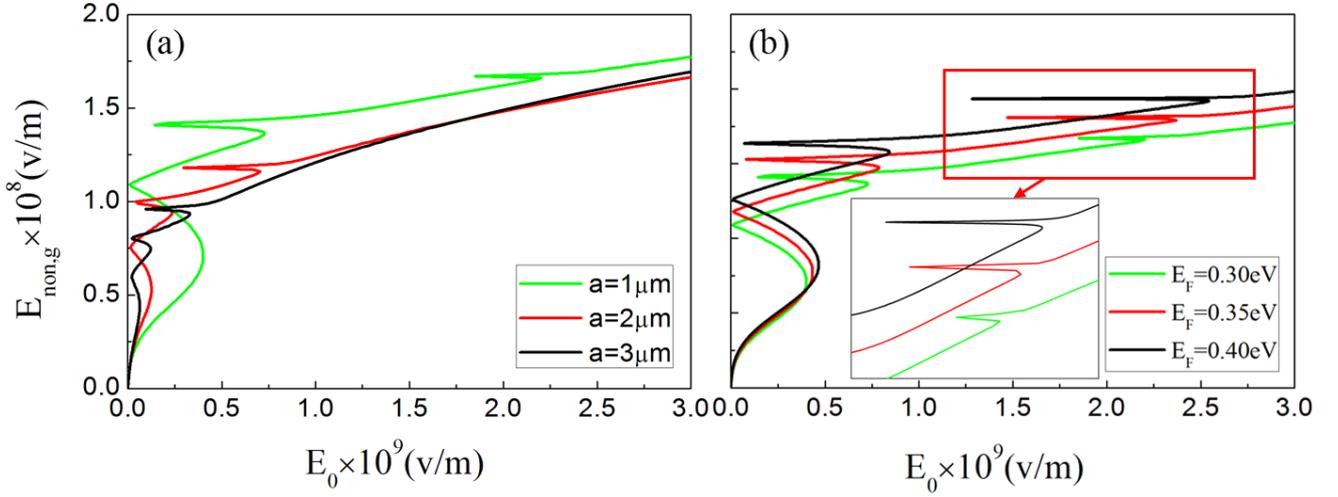

**Fig.7.** Dependence of the average local fields on the applied field for (*a*) **different sizes with** $E_F = 0.3\text{eV}$; and (*b*) **different Fermi energy with** $a = 1\mu m$. Other parameters are $\varepsilon = \varepsilon_h = 2.25$, $\tau = 0.1\text{ps}$ and $\lambda = 40\mu m$.

In the end, we investigate the influence of the particle size and Fermi energy on the multi-stable curves in the near-field. As shown in Fig. 7 (a), the multi-stable region is found to be strongly dependent on the size of nanoparticles, and it is possible to achieve low switching threshold with larger size. In contrast, one might obtain broader OT and OB regions when the Fermi energy is increased. For further increasing the radius of sphere, the octupole modes shifts to the dipole and quadrupole region, resulting in optical multi-stability. Therefore, this graphene-wrapped nanoparticle can provide more freedoms to control its multi-state optical switching, therefore has potential applications in optical communications and computing [34].

## 4. Conclusion

In this paper, we proposed a nonlinear nanoparticle consisting of a dielectric core wrapped by a monolayer graphene layer. Based on nonlinear full-wave theory and self-consistent mean-field method, we study nonlinear optical bistable behavior for the near-field and far-field scattering efficiency in such coated nanoparticle system at terahertz frequencies. We found enhanced local fields near the surface plasmon resonances which could lead to strong nonlinearity. Optical bistability is simultaneously found in curves between the local field, nonlinear scattering efficiency and the external applied field. It is shown that the switching thresholds are highly dependent on the Fermi energy of graphene besides the particle size, therefore, it provides a new degree of freedom to control the local field and scattering field with the input one. Moreover, when the higher-order terms are considered under applying high external electromagnetic field condition, we found multi-bistability in its near field spectra. In details, tristability occurs once we further take into account the quadrupole term, and there exists another bistable region under higher input field if the octupole term is considered. Finally, we investigate the influence of the particle size and Fermi energy on the multi-stable curves. All these novel properties have great potential for the design in optoelectronic devices, like two/three state switching, memory access applications and optical transistor.

## Acknowledgments

This work was supported by the National Natural Science Foundation of China (Grant No. 11374223), the National

Science of Jiangsu Province (Grant No. BK20161210), the Qing Lan project, "333" project (Grant No. BRA2015353), and PAPD of Jiangsu Higher Education Institutions.